\documentclass[twocolumn,notitlepage, superscriptaddress, nofootinbib,nobibnotes, aps,prd,10pt]{revtex4-1}
\usepackage{amsmath,amssymb, fp}
\usepackage{braket}
\usepackage{xcolor}  
\usepackage{soul} 
\usepackage{booktabs}
\definecolor{linkcolor}{RGB}{7,94,84}  
\usepackage[	colorlinks=true,    	
			pdfstartview=FitV,
			linkcolor= linkcolor,
			citecolor= linkcolor,
			urlcolor= linkcolor,
			linktoc=page,
			hyperindex=true,
			hyperfigures=true,  
			breaklinks=true]
			{hyperref}
\usepackage{dblfloatfix}    
\usepackage{balance} 
\usepackage{lastpage}  
\usepackage{mathtools}
\usepackage{epsfig,rotating,pifont}
\usepackage{graphicx}
\graphicspath{{./figures/}}	
\usepackage[caption=false]{subfig}
\usepackage{placeins}

\usepackage{ragged2e} 
\usepackage{etoolbox}
\usepackage{changepage}   
\usepackage{pgfplots}
\usetikzlibrary{pgfplots.groupplots}
\pgfplotsset{compat=1.3}
\usepackage{tikz}  
\usetikzlibrary{arrows,shapes,positioning}
\usetikzlibrary{decorations.markings,decorations.pathmorphing,decorations.pathreplacing,decorations.text}
\usetikzlibrary{arrows,calc,patterns,shapes.geometric}
\usepackage{fancybox}
\usepackage{verbatim}
\usepackage{multirow}
\usepackage{titlesec}

\newsavebox\CBox

\def\beq{\begin{eqnarray}}
\def\eeq{\end{eqnarray}}

\def\la{\langle }
\def\ra{\rangle }

\def\lb{\label}

\def\O{\mathcal{O}}

\newcommand{\Tr}{\mathrm{Tr}\hspace{1pt}}            

\newcommand{\br}{\mathbf{r}}
\newcommand{\bs}{\mathbf{s}}

\def\bea#1\eea{\begin{align}#1\end{align}}
\newcommand{\bg}{\begin{gather}}

\newcommand{\bseq}{\begin{subequations}}
\newcommand{\eseq}{\end{subequations}}

\def\tr{{\rm Tr}}

\def\lb{\label}

\def\nn{\nonumber}

\def\p{\partial}
\def\leftt{\hspace{-2pt}\left(\hspace{-1pt}}
\def\rightt{\hspace{-1.5pt}\right)\hspace{-1pt}}

\def\bes#1\ees{\begin{equation}\begin{split}#1\end{split}\end{equation}}

\definecolor{Green}{RGB}{147,162,153}
\definecolor{Green2}{RGB}{26,148,49}
\definecolor{BrownL}{RGB}{173,143,103}
\definecolor{Red}{RGB}{210,83,60}
\definecolor{BrownD}{RGB}{114,96,86}
\definecolor{GreyD}{RGB}{76,90,106}
\definecolor{GreyB}{RGB}{128,141,160}
\definecolor{Maroon}{RGB}{121,70,61}
\definecolor{Blue}{RGB}{148,184,210}
\definecolor{Blue2}{RGB}{108,144,170}
\definecolor{Blue3}{RGB}{42, 107, 172}
\definecolor{BB}{RGB}{128,184,220}  
\definecolor{teal}{RGB}{7,94,84}  

\newsavebox\foobox
\begin{document}

\title{Full-counting statistics of corner charge fluctuations}  

\author{Cl\'ement Berthiere}
\affiliation{Universit\'{e} de Montr\'{e}al, Département de Physique, Montr\'eal, QC, Canada, H3C 3J7}
\affiliation{Centre de Recherches Math\'{e}matiques, Universit\'{e} de Montr\'{e}al, Montr\'{e}al, QC, Canada, H3C 3J7}

\author{Benoit Estienne}
\affiliation{Sorbonne Université, CNRS, Laboratoire de Physique Théorique et Hautes Energies, LPTHE, F-75005 Paris, France}

\author{Jean-Marie St\'ephan}
\affiliation{Univ Lyon, CNRS, Universit\'e Claude Bernard Lyon 1,\\ Institut Camille Jordan, UMR5208, F-69622 Villeurbanne, France}

\author{William Witczak-Krempa}
\affiliation{Universit\'{e} de Montr\'{e}al, Département de Physique, Montr\'eal, QC, Canada, H3C 3J7}
\affiliation{Centre de Recherches Math\'{e}matiques, Universit\'{e} de Montr\'{e}al, Montr\'{e}al, QC, Canada, H3C 3J7}
\affiliation{\it Institut Courtois, Universit\'e de Montr\'eal, Montr\'eal, QC H2V 0B3, Canada}

\date{\today}

\begin{abstract}\vspace{1pt}
%

Outcomes of measurements are characterized by an infinite family of generalized uncertainties, or cumulants, which provide information beyond the mean and variance of the observable. Here, we investigate the cumulants of a conserved charge in a subregion with corners. We derive nonperturbative relations for the area law, and more interestingly, the angle dependence, showing how it is determined by geometric moments of the correlation function. These hold for translation invariant systems under great generality, including strongly interacting ones. 
We test our findings by using two-dimensional topological quantum Hall states of bosons and fermions at both integer and fractional fillings. We find that the odd cumulants' shape dependence differs from the even ones. For instance, the third cumulant shows nearly universal behavior for integer and fractional Laughlin Hall states in the lowest Landau level. Furthermore, we examine the relation between even cumulants and the R\'enyi entanglement entropy, where we use new results for the fractional state at filling 1/3 to compare these quantities in the strongly interacting regime. We discuss the implications of these findings for other systems, including gapless Dirac fermions, and more general conformal field theories. 

\end{abstract}

\maketitle

\makeatletter
\def\l@subsubsection#1#2{}
\makeatother

A fundamental concept of quantum mechanics is the statistical nature of measurements of observables. Measuring the same observable in identically prepared systems generally leads to different outcomes governed by a probability distribution. This distribution is then described by its cumulants. The higher the order of the cumulant  one has access to, the more detailed information one obtains about correlations in the system. 

In many experiments, one measures a subregion of a physical system only. One is then interested in the cumulants of a given observable in a subregion of the full system.
The cumulants $C_m(A)$ of an observable $\O$ within a subregion $A$ are defined as 
\bea\label{eq:cumu_fcs}
C_m(A)= \p_\lambda^m\log\la e^{\lambda \O_A}\ra|_{\lambda=0}\,.
\eea
The first cumulant is the mean $\la \O_A \ra$. The second and third cumulants, $\la (\O_A -\la \O_A \ra)^m \ra$ with $m=2,3$, are the variance (or fluctuations) and the skewness, respectively. 
Higher-order cumulants are more complicated polynomials in the moments. For instance, the fourth cumulant is given by 
$C_4(A)=\la (\O_A -\la \O_A \ra)^4\ra-3(\la (\O_A -\la \O_A \ra)^2 \ra)^2$.
A nonzero skewness or higher-order cumulant is a signal of the non-Gaussianity of a distribution, since all cumulants of order three and above vanish for a Gaussian one.

As long as the total quantity $\O = \O_A + \O_B$ does not fluctuate, one has $C_m(A) = (-1)^m C_m(B)$ for all cumulants except the mean $m=1$. That is (odd) even cumulants are (anti) symmetric under exchange of the subregion $A$ and its complement $B$. 
In the context of condensed matter, even cumulants of conserved charges---in particular the variance---have received considerable attention \cite{2006PhRvA..74c2306K,Klich:2008un,2010PhRvB..82a2405S,PhysRevB.83.161408,Song:2011gv,Swingle:2011np,2012EL.....9820003C,2012PhRvL.108k6401R,2014JSMTE..10..005P,2019PhRvB..99g5133H,PhysRevB.101.235169,2021PhRvB.103w5108C}. Indeed, some of their properties, are reminiscent of those of entanglement entropy. As it turns out, for non-interacting fermions, 
the full entanglement spectrum is entirely encoded in the \emph{even} charge cumulants~\cite{Klich:2008un,2010PhRvB..82a2405S,PhysRevB.83.161408,Song:2011gv}. Moreover, bipartite charge fluctuations can be shown to be proportional to the (Rényi) entanglement entropy for, e.g., conformal field theories (CFTs) in 1D or Fermi gases \cite{Klich:2008un,2012EL.....9820003C}.
Bipartite fluctuations (and few higher-order charge cumulants) have been measured in mesoscopic condensed-matter systems \cite{2009RvMP...81.1665E,2011PhRvB..83g5432K} and in cold atomic gases \cite{2006NatPh...2..705G,2008RvMP...80..885B}.
The success of entanglement entropy being widespread, relating it to measurable properties has long been the main motivation for studying bipartite fluctuations. More recently, fluctuations have been investigated for topological states and quantum critical systems~\cite{2019PhRvB..99g5133H,Wang:2021lmb,2021ScPP...11...33W,Estienne:2021hbx,Oblak:2021nbj}, as well as in the context of random matrix theory \cite{2019PhRvA..99b1602L,2019PhRvE.100a2137L}.

\medskip\smallskip
\noindent\mbox{\textbf{\textit{General considerations: area law \& corrections.}}}
Cumulants of conserved observables in a pure state behave for large two-dimensional regions $A$ as
\bea\label{eq:cumu_scaling}
C_m(A)= c_{m}\vert\partial A\vert - a_{m} + \cdots\,, \quad 
\eea
where the leading term scales with the area of the boundary of $A$, and $a_m$ is a subleading correction (the minus sign is introduced to match with existing literature on corner terms \cite{Casini:2006hu,2014JSMTE..06..009K,Bueno:2015rda,Berthiere:2018ouo,Estienne:2021hbx,Berthiere:2021nkv}). 
Such an expansion with a leading area law holds in considerable generality: it is satisfied in any translation invariant interacting system, under mild assumptions on the decay of connected correlation functions of the associated local charges, as we show from first principles in Appendix~\ref{sec:ultragenereral}.

An area law is expected for even cumulants from the symmetry between subregion $A$ and its complement $B$: we consider a charge that is globally conserved in the system, such as the number of particles, but it is still allowed to fluctuate between subregions $A$ and $B$. Thus $A$ can trade particles with $B$ through their common boundary, hence one expects an area law. 
Odd cumulants ($m>1$) of conserved observables have been less studied but are also interesting. They are antisymmetric under $A\leftrightarrow B$, which excludes a volume law. Furthermore, the area-law term vanishes for translation and inversion invariant interacting systems, see again Appendix~\ref{sec:ultragenereral}.
Thus neither volume nor area-law terms appear, that is $c_m=0$ for $m$ odd in \eqref{eq:cumu_scaling}, and the leading contribution is $ - a_{m}$. 

The (subleading) piece $a_m$ in expansion \eqref{eq:cumu_scaling} is the most interesting, as it is sensitive to the presence of corners in $A$. (If the boundary of $A$ is smooth, this term vanishes, and other contributions appear from the curvature of the boundary \cite{BESW_curvature}, but those enter at an order proportional to the inverse size of $A$.) This corner contribution depends on the opening angle $\theta\in (0,2\pi)$; if there are several corners, $a_m$ is obtained by summing over all corner contributions, and those are independent. In practice, isolating the contribution of a \emph{single corner} can be challenging.
Hereinafter, we denote by $a_m(\theta)$ the contribution of a single corner of opening angle $\theta$.

Corner cumulant functions exhibit universal behavior. We  show non-perturbatively that in the cusp limit $\theta\rightarrow0$, they all diverge as
\bea\label{cusplim}
\qquad a_m(\theta) \sim \kappa_m/\theta \,,\qquad( \theta \to 0)\,.
\eea
This is done by considering a cumulant on a general parallelogram and identifying the leading term as two of the angles go to zero (Appendix \ref{sec:smallangles}). The cusp coefficient $\kappa_m$ is given by the following geometric moment of the $m$--point connected correlation function $f$,
\bea\lb{kappa_correl}
\hspace{-4pt}\kappa_m=-\frac{1}{2}\!\int_\mathbb{C} d z_2\ldots d z_{m}\,  M[y_2,\ldots,y_{m}]^2 f(z_2,\ldots,z_m)\hspace{1pt},\hspace{-2pt}
\eea
where we used complex coordinates $z_j=x_j + i y_j$. 
$M[y_2,\ldots,y_{m}]$ is defined in the Appendix, see (\ref{eq:maxmin}), and reduces to $M[y_2]=|y_2|$ for the variance while $M[y_2,y_3]=(|y_2|+|y_3|+|y_2-y_3|)/2$ for the third cumulant.
The behavior of the corner contribution in the cusp limit is thus universal, valid for any translation invariant system provided $f$ does not decay too slowly. 

Another property common to all corner cumulant functions $a_m(\theta)$ is that they must vanish at $\theta=\pi$ since the corner disappears. In the smooth regime $\theta\sim \pi$, we must distinguish even and odd cumulants. Because of the invariance of even cumulants under the exchange of region $A$ with its complement, we have $a_m(\theta)=a_m(2\pi-\theta)$. This implies that even $a_m(\theta)$ vanish quadratically in the smooth limit $\theta\sim \pi$, which is non-singular,
\bea \label{eq:smooth-even}
\qquad\quad a_m(\theta\sim\pi) \simeq \sigma_m\cdot (\pi-\theta)^2 \,, \qquad m\;\,{\rm even}\,.
\eea
In contrast, odd cumulants are antisymmetric under this exchange, implying $a_m(\theta)\hspace{-.5pt}=\hspace{-.5pt}-a_m(2\pi-\theta)$, so 
we expect $a_m(\theta)$ to vanish linearly in the smooth regime,
\bea\lb{eq:smooth-odd}
\qquad\quad a_m(\theta\sim\pi)\simeq \sigma_m \cdot (\pi-\theta)\,, \qquad m\;\,{\rm odd}\,.
\eea
The smooth coefficient $\sigma_m$ is expected to be related to certain sum rules and to encode universal properties of the system, as is the case for the variance \cite{Estienne:2021hbx}.

In what follows, we systematically study higher-order cumulants which are much more complicated, focusing on the example of quantum Hall states.  
We present results for the charge cumulants of the IQH groundstate at filling $\nu=1$. The cumulants can be computed using standard free fermion methods~\cite{2003JPhA...36L.205P,Klich:2004pb}. We also consider Laughlin states at filling fractions $\nu=1/2$ (bosons), $\nu=1/3$ (fermions) where fluctuations are accessible through Monte Carlo simulations.

\begin{figure}[t]
\vspace{.2cm}
\centering
\includegraphics[scale=0.63]{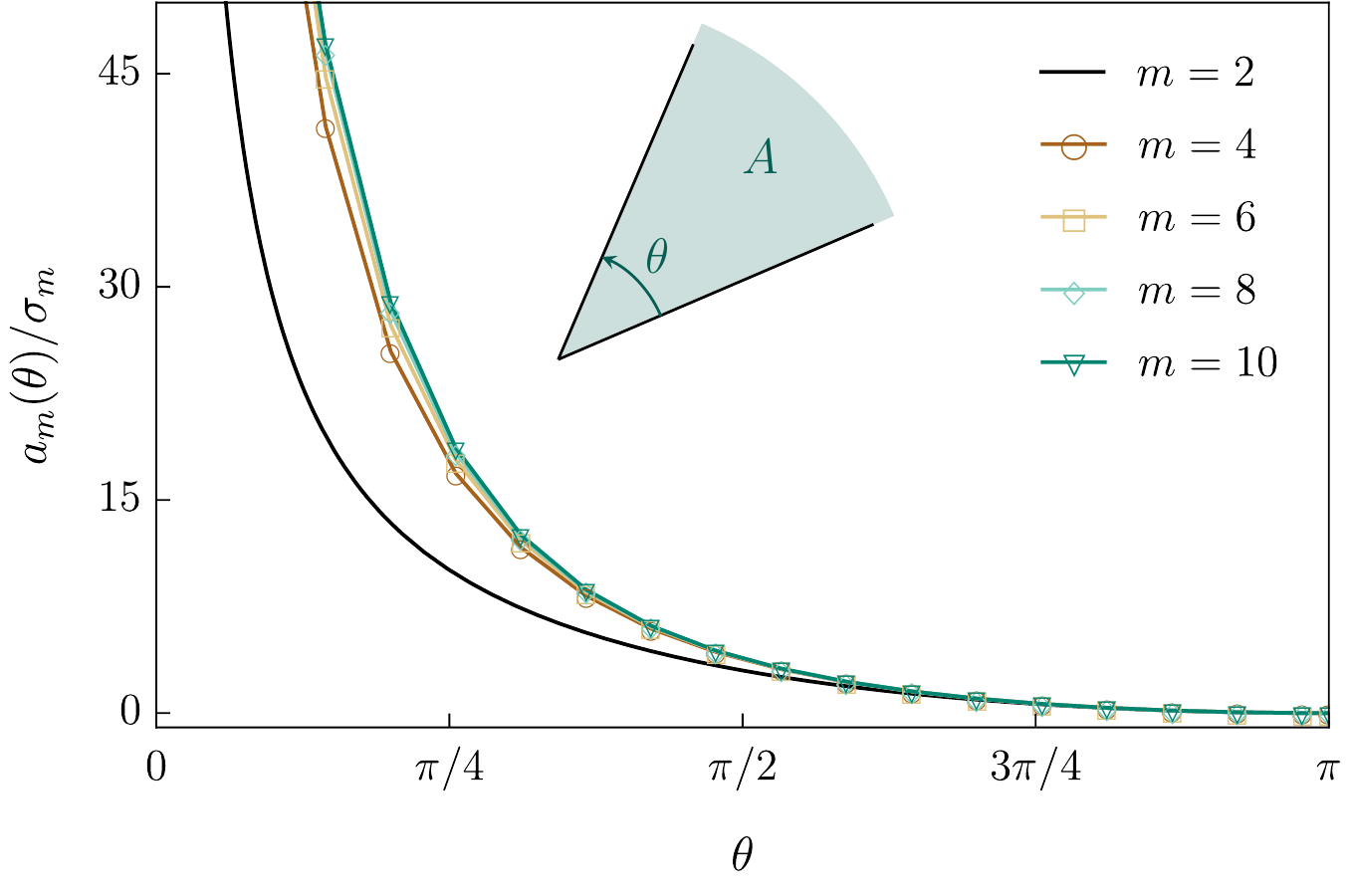}
\vspace{-15pt}
\caption{Normalized corner cumulant function $a_m(\theta)/\sigma_m$ for $m=2,4,\cdots,10$ for IQH state at filling $\nu=1$. The variance stands out while the higher cumulants cluster. Embedded is shown a region $A$ with a corner of opening angle $\theta$.}
\lb{fig_even}
\end{figure}

\bigskip
\noindent\textit{\textbf{Even cumulants in quantum Hall states.}} 
For large regions, the cumulants behave as (\ref{eq:cumu_scaling}).
All $c_m$ are known exactly for IQH \cite{2019CMaPh.376..521C}. 
For example, the second cumulant has area-law coefficient \cite{Klich:2004pb} $c_2=(2\pi)^{-3/2}$ for $\nu=1$ (working in units of the magnetic length $\ell_B=1$), while the corner fluctuations function for general incompressible filling $\nu$ reads \cite{Estienne:2021hbx}:
$a_2(\theta) = \frac{\nu}{4\pi^2}\big(1+(\pi-\theta)\cot\theta\big)$.
Both coefficients $c_m$, $a_m$ can be numerically computed to high precision using the method of \cite{2010JSMTE..12..033R}; details and values of $c_m$ for $m=2,4,\cdots,28$ in Appendix~\ref{sec:exactnumerics} and Fig.\,\ref{SM:fig1}. We find that the area-law coefficients are not always positive, but rather alternate in sign as $(-1)^{m/2-1}$. The fourth cumulant is thus negative, the sixth positive, and so on. The coefficient $c_m$ takes its minimal value at $m=8$, and then increases factorially with the order $m$. This factorial growth is expected from the definition of the cumulants, see \eqref{eq:cumu_fcs} and \eqref{eq:fcs_area}.

The corner term $a_m(\theta)$ behaves similarly. For fixed $\theta$, as the order $m$ increases, $a_m(\theta)$ first decreases to its minimal value and then increases factorially. However, $a_m(\theta)$ changes sign differently than $c_m$. Indeed, both $a_2$ and $a_4$ are positive functions, while $a_6$ is negative, $a_8$ positive, and the signs continue to alternate. 
As a consequence, in all the cases that we studied, i.e.~$m\leq 28$, only the second cumulant presents an area-law term of opposite sign compared to its subleading corner contribution. 
Starting with the fourth cumulant, area-law and subleading corner terms have same sign. 
For $m>2$, the coefficients $a_m$ and $c_m$ probe somewhat complicated sum rules for the connected $m$--point density function, as is explained in Appendix~\ref{sec:ultragenereral}. The sign of those sum rules cannot be easily understood from underlying general physical principles.

\begin{table}[t]
\vspace{-4pt}
\caption{\label{tab1}Smooth and cusp coefficients for $\nu\!=\!1$ cumulants.}
\begin{ruledtabular}
\begin{tabular}{ l r r }
 & $\sigma_m\quad\,$ & $\kappa_m\;\;$ \vspace{3pt}\\
\colrule
$C_2$ \vspace{-1pt}& $1/(12\pi^2)$ & $1/(4\pi)\simeq 0.07957$ \\
$C_4$ \vspace{-1pt}& $0.000364$ & $\frac{18+\pi-12\sqrt{3}}{4\pi^2}\simeq 0.00904$ \\
$C_6$ \vspace{-1pt}& $-0.000191$ & $-0.00459$ \\
$C_8$ \vspace{-1pt}& $0.000221$ & $0.00478$ \vspace{2pt}\\
\colrule
$C_3$ \vspace{-1pt}& $-0.01462$\phantom{0} & $-\frac{3\sqrt{3}-\pi}{4\pi^2}\simeq -0.05204$ \\
$C_5$ \vspace{-1pt}& $0.00254$\phantom{0} & $0.01168$ \\
$C_7$ \vspace{-1pt}& $-0.00171$\phantom{0} & $-0.00938$ \\
\end{tabular}
\end{ruledtabular}
\end{table}

\begin{table}[b]\vspace{-5pt}
\caption{\label{tab2}Area-law coefficients $c_m$, $m=2,4$, for the IQH $\nu=1$ state (exact), and the FQH ones at $\nu=1/2,\,1/3$ (MC).}
\begin{ruledtabular}
\begin{tabular}{ l r r r }
& $\nu=1\hspace{8pt}$ & $\nu=1/2\hspace{2pt}$ & $\nu=1/3$ \vspace{3pt}\\
\colrule
$c_2$ \vspace{-1pt}& $(2\pi)^{-3/2}\simeq0.0635$ & $0.0351$ & $0.0255$ \\
$c_4$ \vspace{-1pt}& $-0.00336$ & $-0.00219$ & $-0.00163$ \\
\end{tabular}
\end{ruledtabular}
\end{table}

Corner cumulant functions $a_m(\theta)$ share important features, such as their behavior in the cusp \eqref{cusplim} and smooth \eqref{eq:smooth-even} regimes. We expect even $a_m(\theta)$ to be monotonic functions for $0<\theta<\pi$, as exemplified in Fig.\,\ref{fig_even} for quantum Hall states.
Both properties \eqref{cusplim} and \eqref{eq:smooth-even} can be explicitly verified for the variance, and also hold for the R\'enyi entropies at both integer \cite{Sirois:2020zvc} and fractional fillings~\cite{Estienne:2021hbx}, as well as in 2D CFTs.
The values of the smooth $\sigma_m$ and cusp $\kappa_m$ coefficients for $m=2,4,\cdots,8$ may be found in Table~\ref{tab1}. Remarkably, we were able to find analytical expressions for $\kappa_m$, in terms of $m$--fold integrals (see Appendix~\ref{sec:smallangles}). Those integrals simplify nicely for $m=2,3,4$, as reported in Table~\ref{tab1}. 
The smooth coefficient has been used to normalize the corner cumulant functions in Fig.\,\ref{fig_even},
where we notice that the second cumulant stands out from the higher-order ones. It would be interesting to determine whether this dichotomy, and the clustering of higher-cumulants, hold in other quantum states.

\begin{figure}[t]
\vspace{.2cm}
\centering
\includegraphics[scale=0.684]{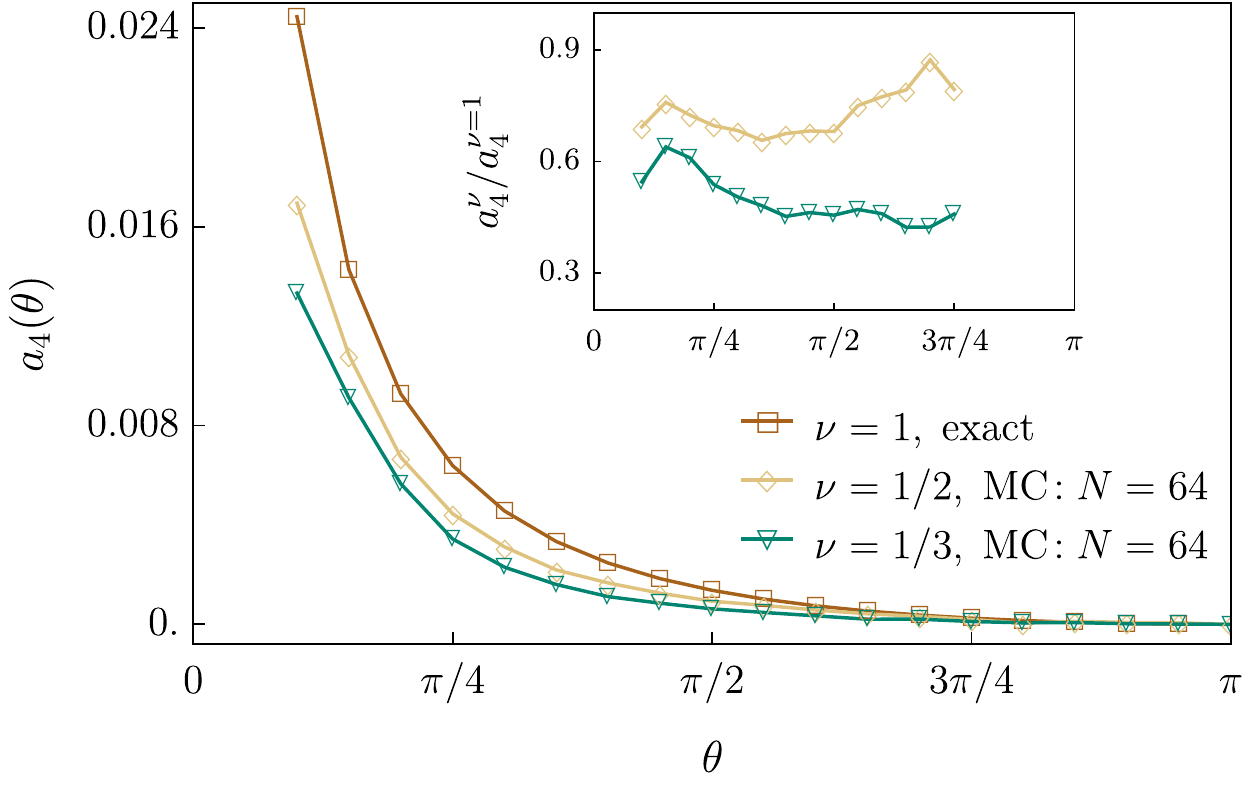}
\vspace{-15pt}
\caption{Fourth corner cumulant function for the Laughlin state at fillings $\nu=1,1/2,1/3$, extracted from Monte Carlo data on $N=64$ particles. The inset shows the ratio $a^\nu_4(\theta)/a^{\nu=1}_4(\theta)$ for $\nu=1/2,1/3$.}
\lb{fig_a4}
\end{figure}

\medskip
\noindent\textit{Fractional Quantum Hall.---} We now study fractional quantum Hall (FQH) states, focusing on Laughlin states in a disc geometry at fillings $\nu=1/2$ (bosons) and $1/3$ (fermions), using large-scale Monte Carlo (MC) simulations. Computing cumulants is done by counting the number of particles in $A$ for each Monte Carlo sample. Extracting the corner contribution requires more work, and was done using the same method as in \cite{2020PhRvB.101k5136E,Estienne:2021hbx}, which focused on the variance and second Rényi entropy. 

For fractions $\nu=1/2, 1/3$, we found that the fourth cumulant is negative, i.e.~$c_4<0$ as for the IQH state at $\nu=1$, see Table~\ref{tab2}. We show in Fig.\,\ref{fig_a4} the corresponding corner function $a_4(\theta)$, which is positive for both FQH states, exactly as we found for $\nu=1$. The angular dependence for fillings $\nu=1,1/2,1/3$, although similar, is not superuniversal, in contrast with fluctuations where $a_2^{\nu_1}(\theta)/a_2^{\nu_2}(\theta) = \nu_1/\nu_2$ (the ratios are plotted in the inset of Fig.\,\ref{fig_a4}).
This breakdown of superuniversality is expected. Intuitively, the constant term comes from the neighborhood of the corner, which looks like an infinite angular sector $A=\{z,0\leq \arg z\leq \theta\}$ in 2D complex coordinates. The second cumulant involves the two-point translationally invariant density function $\braket{\rho(0)\rho(z)}$, and superuniversality essentially follows from  scale invariance of the region $A$, see \cite{Estienne:2021hbx}. For higher cumulants, scale invariance is no longer sufficient to constrain the angular dependence of the corner term, because higher translationally invariant correlation functions depend on more than one variable. 
An analytical counterexample to superuniversality for $C_3$ is given in Appendix~\ref{sec:smallangles}, which relies on the two special points $\theta\to 0$ and $\theta=\pi/2$.

We observe that $|C_4|$ increases with the filling fraction, which is in agreement with the intuition that having a greater density of particles leads to greater particle fluctuations in $A$. In fact, the increase holds for both the area-law coefficient and $a_4(\theta)$.
The same holds for the variance $C_2$, where the corner term scales exactly linearly with $\nu$.
Interestingly, for filling fractions $\nu=1/3,\hspace{1pt}1/2,\hspace{1pt}1$,
we found that the area-law coefficient $c_2$ also depends linearly (within $0.2\%$ relative error) on $\nu$: $c_2(\nu)\simeq 0.00653 + 0.05699\hspace{1pt}\nu$, see Table~\ref{tab2}. No simple explanation for this observed linearity is known since $c_2$ depends on the entire static structure factor, not only its long-distance part~\cite{Estienne:2021hbx}.  For the fourth cumulant, $c_4$ is very close to such a linear behavior as well. We note that at integer fillings $\nu >1$, the area-law coefficient $c_2$ increases with $\nu$, but sublinearly (see Fig.\,\ref{SM:fig1} in Appendix \ref{sec:exactnumerics}). There is thus a striking difference in the behavior of $c_2$ between fillings $\nu \leq 1$ and integer ones $\nu>1$.

\medskip

\noindent\textit{From cumulants to entanglement.---} For systems that map to free fermions with conserved particle number, the full counting statistics associated with the bipartite charge fluctuations encodes the full entanglement spectrum \cite{2006PhRvA..74c2306K,Klich:2008un,PhysRevB.83.161408}. The R\'enyi entanglement entropies are determined by the full set of the \textit{even} charge cumulants, which can be understood from the fact that entropies are symmetric between $A$ and its complement $B$ for pure states. Using the series representation \cite{PhysRevB.83.161408} of the entanglement entropy in terms of the even cumulants with increasing cutoff number $M$, we observe that the corner contribution converges slowly (as $\propto M^{-1}$) but monotonically to the exact answer, from below (in absolute value). This monotonicity is nontrivial since the even corner cumulant functions alternate in sign whereas the series coefficients do not, although cumulants higher than the variance are rapidly suppressed in the series.

There is no such relation between charge cumulants and R\'enyi entropies for interacting systems \cite{2014JSMTE..10..005P}. We have computed the corner contribution to the second R\'enyi entropy for the Laughlin state at fractional $\nu=1/3$ using the method of \cite{Estienne:2021hbx}. Applying the formula for noninteracting systems with our results for the second and fourth charge cumulants at $\theta=\pi/2$, we find $0.0195$, which is less than the R\'enyi corner term $ 0.1238$.  
For the IQH state at $\nu=1$, the estimate using the first two even cumulants, $0.0586477$, is also lesser than the true R\'enyi entropy 
$0.0617735$ \cite{Sirois:2020zvc}, but the difference is much smaller than for FQH. Whether a relation between entanglement entropies and cumulants, such as a bound, could be esta-blished for interacting systems is an important question.

\bigskip
\noindent\textit{\textbf{Odd cumulants in quantum Hall states.}} 
As already mentioned, odd cumulants do not scale with the area of the boundary of $A$. 
For regions with corners, $a_m(\theta)$ is the leading term. The data confirm the expected behavior of $a_m(\theta)$ in the cusp \eqref{cusplim} and smooth \eqref{eq:smooth-odd} regimes, see Fig.~\ref{fig_a3}. As for even cumulants, we observe a pattern of alternating sign of $a_m(\theta)$ as $(-1)^{(m-1)/2}$ in the range $0<\theta\le\pi$ for the IQH data (see Appendices).
The odd corner cumulant functions all appear to be monotonic, an observation which is not unreasonable to expect.

\medskip
\noindent\textit{Fractional Quantum Hall.---} We performed Monte Carlo  simulations on the third cumulant for the Laughlin states 
at filling fractions $\nu=1/2,1/3$. In Fig.\,\ref{fig_a3}, we show our results for $C_3(\theta)$, found positive for both fractional states as for the IQH state at unit filling.  
We observe that $C_3$ increases with the filling fraction, similarly as $C_2$ and $|C_4|$.
Quite remarkably, the data suggest that the angular dependence of $C_3$ is universal for filling fractions $\nu=1/3,\hspace{1pt}1/2,\hspace{1pt}1$, as can be seen from the the inset of Fig.\,\ref{fig_a3} where we have plotted the ratio $a^\nu_3(\theta)/a^{\nu=1}_3(\theta)$, which is constant over a wide range of angles. We find $a^\nu_3(\theta)/a^{\nu=1}_3(\theta)\simeq0.67$ and $0.53$ for $\nu=1/2$ and $\nu=1/3$, respectively. Furthermore, the dependence on $\nu$ is nearly linear:
$a_3^\nu(\theta)/ a_3^1(\theta) \simeq \alpha\nu +\beta$,
where $\alpha=7/10=1-\beta$ (within $3\%$ error relative to the fit in Fig.\,\ref{fig_a3}).
It is striking that $C_3$ displays such universality for the three states with $\nu\leq 1$ given their very different properties.

\begin{figure}[t]
\vspace{2pt}
\centering
\includegraphics[scale=0.685]{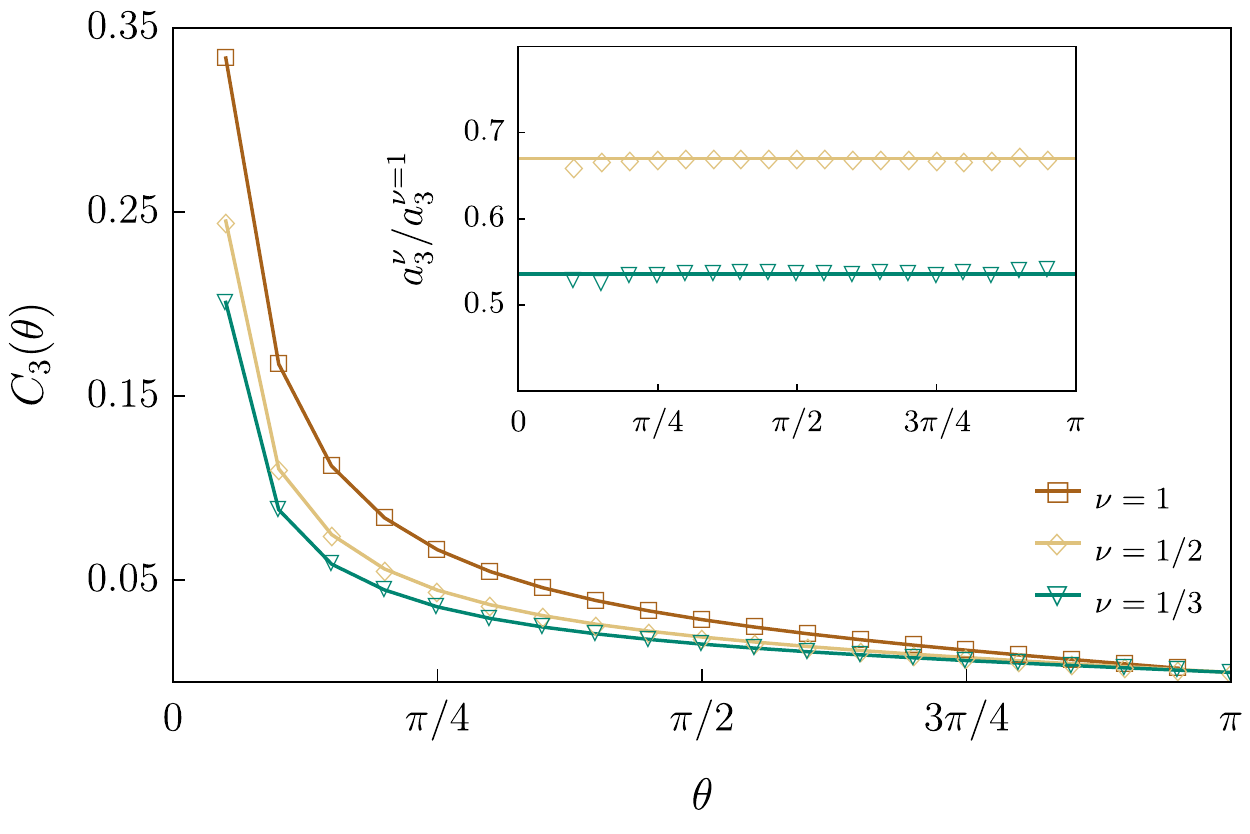}
\vspace{-15pt}
\caption{Third corner cumulant for IQH $\nu=1$ (exact) and FQH $\nu=1/2,1/3$ (MC on $N=64$ particles). The inset shows the ratio $a^\nu_3(\theta)/a^{\nu=1}_3(\theta)$ for $\nu=1/2,1/3$.}
\lb{fig_a3}
\vspace{-5pt}
\end{figure}

\bigskip
\noindent\textit{\textbf{Discussion.}} 
We have studied the cumulants of a conserved charge in subregions with corners. The (subleading) contribution $a_m(\theta)$ is sensitive to finer geometric details of the subregion such as corners. We have derived a nonperturbative relation for the angle dependence, see \eqref{cusplim} and \eqref{kappa_correl}, showing how it is determined by geometric moments of the correlation function. This hold for translation invariant systems under great generality, including strongly interacting ones. We also expect the behavior of $a_m(\theta)$ in the smooth regime to hold in considerable generality, as well as their monotonicity for $0<\theta\le\pi$. We have then tested our findings with 2D quantum Hall states at both integer and fractional fillings.

The variance is known to be superuniversal \cite{Estienne:2021hbx,BESW_curvature}, i.e.~it takes the same form for a large class of unrelated systems. We have shown that superuniversality breaks down for cumulants higher than the variance. However, we have discovered that the angle dependence of the third cumulant appears to be universal within error bars over a wide range of angles for quantum Hall states at fillings $\nu=1,1/2,1/3$. 
These numerical results give access to information on sum rules for higher correlation functions, but in a more convoluted way than for the variance. Studying such subregion cumulants provides a new method to understand such non trivial sum rules, in particular the third cumulant which displays striking universality for quantum Hall states at fillings $\nu\le1$.

An interesting direction would be to study charge cumulants in other systems such as CFTs in $d\geq 2$ spatial dimensions, beyond the variance~\cite{Estienne:2021hbx,2021ScPP...11...33W,Wang:2021lmb}. When considering a conserved charge $\mathcal O$, the cumulant-generating function is the expectation value of the so-called disorder operator $\exp(i\lambda \mathcal O_A)$, which performs a symmetry transformation in subregion $A$. The expectation value of the disorder operator, which can be used to probe higher-form symmetries, was studied as a function of both $\lambda$ and the corner angle near quantum critical points~\cite{Zhao:2020vdn,Wang:2021lmb,2021ScPP...11...33W}. However, little is understood for higher cumulants $C_{m>2}$. For conserved currents of CFT, the three-point function vanishes at equal times \cite{Osborn:1993cr}. The corresponding third cumulant thus vanishes as well, in contrast to what we found for quantum Hall states. For CFTs with charge 
conjugation symmetry $\mathcal C$, all odd cumulants vanish as well since the charge density $J_0$ is odd under $\mathcal C$. In fact, this vanishing of odd cumulants is general to $\mathcal C$-symmetric systems, independent of details, since the charge density is always odd. This is for example the case for tight-binding models of hopping electrons that are particle-hole symmetric, see \cite{2021PhRvB.103w5108C} for an example with 2D Dirac cones.   
In contrast, not much can be said for even cumulants beyond the variance, even about the sign. For instance, we considered a tight-binding model on the square lattice with Dirac cones, and found that the sign of the leading area-law coefficient $c_m$ for even charge cumulants is $+,+,-,+,+,-,\cdots$ for $m=2,4,6,8,10,12,\cdots$, displaying a pattern different than the one observed for quantum Hall states ($+,-,+,-,+,-\cdots$; the first two signs hold for the FQH states as well). We have shown this by numerically calculating the cumulants, see Appendix \ref{sec:dirac}. It would be interesting to extend this analysis to higher corner terms in Dirac semimetals and general CFTs, since they encode new universal information about the quantum critical degrees of freedom.

\medskip
\noindent\textit{\textbf{Acknowledgments.}} 
W.W.-K. thanks P.-G.~Rozon for earlier collaboration on related topics. B.E. thanks N.~Regnault for discussions on the effect of charge-conjugation on the spectrum of the correlation matrix. 
C.B. was supported by a CRM-Simons Postdoctoral Fellowship at the Université de Montréal. B.E. was supported by Grant No.~ANR-17-CE30-0013-01. J.-M.S. was supported by IDEX Lyon project ToRe (Contract No.~ANR-16-IDEX-0005). W.W.-K. was funded by a Discovery Grant from NSERC, a Canada Research Chair, and a grant from the Fondation Courtois.


\let\oldaddcontentsline\addcontentsline
\renewcommand{\addcontentsline}[3]{}

\bibliographystyle{utphys} 
\bibliography{biblio_cumulants.bib}

\let\addcontentsline\oldaddcontentsline

\onecolumngrid
\clearpage
\begin{center}
\textbf{\large Supplemental Material: Full-counting statistics of corner charge fluctuations}\\[.45cm]
  Cl\'ement Berthiere$^{1,2}$, Benoit Estienne$^3$, Jean-Marie Stéphan$^4$ \,and William Witczak-Krempa$^{1,2,5}$\\[.15cm]
  {\itshape \small ${}^1$Universit\'{e} de Montr\'{e}al, Département de Physique, Montr\'eal, QC, Canada, H3C 3J7\\
  ${}^2$Centre de Recherches Math\'{e}matiques, Universit\'{e} de Montr\'{e}al, Montr\'{e}al, QC, Canada, H3C 3J7\\
  $^3$Sorbonne Université, CNRS, Laboratoire de Physique Théorique et Hautes Energies, LPTHE, F-75005 Paris, France\\
  $^4$Univ Lyon, CNRS, Université Claude Bernard Lyon 1,\\ Institut Camille Jordan, UMR5208, F-69622 Villeurbanne, France\\
  $^5$Institut Courtois, Universit\'e de Montr\'eal, Montr\'eal, QC H2V 0B3, Canada\\}
{\small (Dated: \today)}\vspace*{-0.55cm}
\end{center}

\setcounter{equation}{0}
\setcounter{figure}{0}
\setcounter{table}{0}
\setcounter{page}{1}
\makeatletter
\renewcommand{\thesection}{\Roman{section}}
\renewcommand{\thesubsection}{\arabic{subsection}}
\renewcommand{\theequation}{S\arabic{equation}}
\renewcommand{\thefigure}{S\arabic{figure}}
\renewcommand{\thetable}{S\Roman{table}}

\addtocontents{toc}{\vspace{-17pt}}
\tableofcontents 

\medskip


\setcounter{section}{0}

\section{Scaling of cumulants from general principles}
\label{sec:ultragenereral}
\noindent In this appendix, we gather several general results regarding the scaling of cumulants in general interacting theories. 

\subsection{Disentangling geometry and correlation functions}
Consider a general interacting system in the continuum, and $A$ a region of $\mathbb{R}^d$. The $m$--th cumulant for fluctuations may be written as
\begin{equation}
    C_m(A)=\int_{A^m}d\br_1\ldots d\br_m  \braket{\rho(\br_1)\ldots \rho(\br_m)}_c  \,,
\end{equation}
where $\rho(\br)$ is the local density associated to the conserved quantity, and $\braket{\rho(\br_1)\ldots \rho(\br_m)}_c$ its connected $m$-point function. In the following, we assume that the theory under consideration is invariant with respect to translations
\begin{equation}
    \braket{\rho(\br_1)\ldots \rho(\br_m)}_c=f(\br_2-\br_1,\ldots,\br_m-\br_1),
\end{equation}
where notice $f$ has $m-1$ variables. We further assume that $f(\bs_2,\ldots,\bs_m)$ decays faster than any power law when any of its argument has large modulus. For example, all bulk quantum Hall states considered in this paper satisfy those two requirements.

Using the first assumption, and following the approach put forward in \cite{Kac1954Toeplitz,Widom_TI} to study a similar problem, the $m$--th cumulant can be rewritten after change of variables as
\begin{equation}\label{eq:funda}
    C_m(A)=\int_{\mathbb{R}^{(m-1)d}}d\bs_2\ldots d\bs_m\, \mathcal{G}_A(\bs_2,\ldots ,\bs_m)f(\bs_2,\ldots,\bs_m)\,,
\end{equation}
where $\mathcal{G}_A$ is defined as
\begin{equation}
    \mathcal{G}_A(\bs_2,\ldots,\bs_m)=\int_{\mathbb{R}^d}d\bs\, \mathbf{1}_{\{\bs\in A\}}\mathbf{1}_{\{\bs_2+\bs\in A\}}\ldots \mathbf{1}_{\{\bs_m+\bs\in A\}} \,.
\end{equation}
Here $\mathbf{1}_{\{c\}}$ evaluates to one if condition $c$ is satisfied, zero otherwise. $\mathcal{G}_A$ is a purely geometric quantity, in fact it is nothing but the volume of the region $A\cap (A-\bs_2) \cap \ldots \cap (A-\bs_m).$

With this at hand and using our second assumption, finding a full asymptotic expansion of the cumulants $C_m(\lambda A)$ as $\lambda\to\infty$ boils down to finding an asymptotic expansion for $\mathcal{G}_{\lambda A}$ as $\lambda\to\infty$. Since
\begin{equation}
    \mathcal{G}_{\lambda A}(\bs_2,\ldots,\bs_m)=\lambda^d \mathcal{G}_A(\bs_2/\lambda,\ldots,\bs_m/\lambda)\,,
\end{equation}
this only requires understanding the expansion of $\mathcal{G}_{A}$ for small arguments. The first term is simply proportional to volume, since by definition $\mathcal{G}_A(0,\ldots,0)=\textrm{vol}\, A$. For regions with a smooth boundary, or for polygons, the next order term is \cite{Widom_TI}
\begin{equation}\label{eq:widom_formula}
    \mathcal{G}_{\lambda A}(\bs_2,\ldots,\bs_m)=\lambda^d \textrm{vol}\,A-\lambda^{d-1} \int_{\partial A} d\sigma \max(0,\bs_2\cdot\mathbf{n}_\sigma,\ldots ,\bs_m\cdot\mathbf{n}_\sigma)+o(\lambda^{d-1})\,,
\end{equation}
where the integral is on the boundary of $A$, $\partial A$, and $\mathbf{n}_\sigma$ is the unit outer normal at a given point of $\partial A$ parameterized by $\sigma$. Higher order corrections have also been studied for smooth boundaries \cite{roccaforte,widom_monograph}. The result takes the form of a full asymptotic series, but explicit expressions for each term become very quickly cumbersome as order increases. For polytopes, the higher order expansion takes a different form, in particular it terminates at order $\lambda^0$, due to the fact that the intersection of translated polytopes is still a polytope. As already alluded to, these expansions for $\mathcal{G}_{\lambda A}$ may then be plugged in (\ref{eq:funda}) to get an asymptotic expansion for the cumulants.
\subsection{Sum rules and asymptotic expansion of the cumulants}
While the general structure of the asymptotic expansion was described above, certain terms might vanish due to the symmetries of the physical model under consideration. For example, particle number conservation imposes the sum rule
\begin{equation}
    \int_{\mathbb{R}^d} d\br_2\, f(\br_2,\ldots,\br_m)=0 \,,
\end{equation}
which means the volume term vanishes for all cumulants in case particle number is conserved. This is the famous area-law scaling for even cumulants. For odd cumulants, one can show by similar symmetry considerations that the area-law term also vanishes provided inversion symmetry $f(-\br_2,\ldots,-\br_m)=f(\br_2,\ldots,\br_m)$ is also present.

\subsection{Examples of exact geometric formulas}

Besides the general asymptotic result (\ref{eq:widom_formula}), there are several geometries for which $\mathcal{G}_A(\br)$, or even $\mathcal{G}_A(\br_2,\ldots,\br_m)$ can be computed in closed form. We discuss two of them below, the circle and the square. Before doing so, let us mention the following ``linear transformation formula'':
\begin{equation}\label{eq:transfo}
    \mathcal{G}_{u(A)}(\br_2,\ldots,\br_m)=(\det u)\,\mathcal{G}_A(u^{-1}(\br_2),\ldots,u^{-1}(\br_m))\,,
\end{equation}
where $u$ is any linear map with strictly positive determinant. The proof of this formula follows from either linearity and change of variables, or the interpretation as volume. 

\medskip
\paragraph{The circle}

\begin{figure}[t]
    \centering
    \includegraphics[scale=1.25]{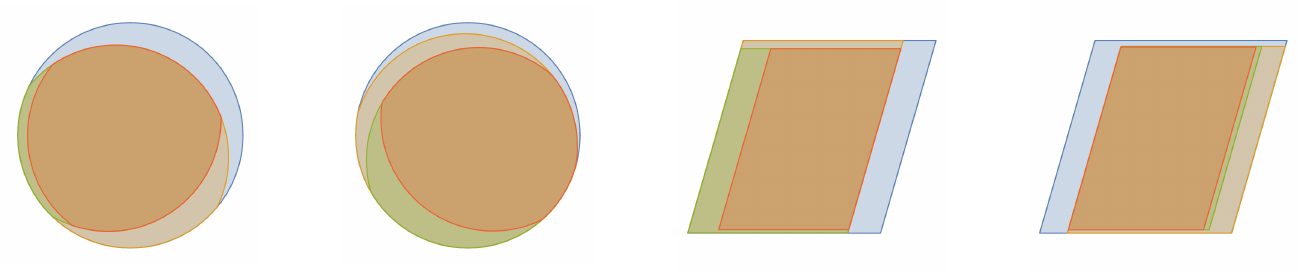}\qquad
    \caption{$A$ (blue), $A\cap (A-\br_2)$ (orange), $A\cap(A-\br_2)\cap(A-\br_3)$ (green), $A\cap(A-\br_2)\cap(A-\br_3)\cap(A-\br_4)$ (red) for several choices of vectors  $\br_2,\br_3,\br_4$. The two left pictures are for $A$ a disk, the two right for $A$ a parallelogram. As can be seen, multiple intersections of translated  parallelograms still give a parallelogram, and one can exploit this to get a simple explicit formula for $\mathcal{G}_A(\br_2,\ldots,\br_m)$ for any $m\geq 2$, see (\ref{eq:para}). The asymptotic regime we are interested in corresponds to the case where all $\br_j$ are small compared to the size of $A$.}
    \label{fig:intersections}
\end{figure}

As suggested by Fig.\,\ref{fig:intersections}, establishing a simple formula for all cumulants seems very complicated, but for the second cumulant one can establish this
\begin{equation}
    \mathcal{G}_A(r)=2R^2 \arccos \frac{r}{2R} -R r \sqrt{1-\left(\frac{r}{2R}\right)^2}\,,
\end{equation}
for a disc of radius $R$. Here $r=|\br|$, and $0\leq r\leq R$. As explained in the previous  subsection, a full asymptotic expansion of the variance is obtained by large $R$ (or small $r$) expansion of the above formula, which reads
\begin{equation}
    \mathcal{G}_A(r)=\pi R^2-R^2\sum_{n\geq 0} \alpha_n \left(\frac{r}{R}\right)^{2n+1}\,,
\end{equation}
for coefficients $\alpha_n$ which can easily be computed. Notice the absence of constant term in the series. 
Generalization to an ellipse can be done using (\ref{eq:transfo}), see \cite{BESW_curvature} for a further discussion.

\medskip
\paragraph{The parallelogram}
For the interval $[0,1]$, one can show that
\begin{equation}\label{eq:exact1d}
    \mathcal{G}_{[0,1]}(x_1,\ldots,x_n)=1-M[x_1,\ldots,x_n]\,,
\end{equation}
where $M$ is defined as
\begin{equation}\label{eq:maxmin}
    M[x_1,\ldots,x_n]=\max(0,x_1,\ldots,x_n)-\min(0,x_1,\ldots,x_n) \,,
\end{equation}
see, e.g., \cite{Kac1954Toeplitz}.
Equation (\ref{eq:exact1d}) holds provided $M[x_1,\ldots,x_n]\leq 1$---which is always true in any relevant asymptotic regime---otherwise $\mathcal{G}_{[0,1]}=0$. Using this result one can deduce the analog formula for a square, and then the parallelogram by using the linear transform formula (\ref{eq:transfo}). For example, if $A=u(S)$ is the image of the square $S=[0,L]^2$ through the linear map $u((x,y))=(x+a y,y)$, then $u(S)$ is a parallelogram with angles $\theta,\pi- \theta,\theta,\pi-\theta$, where $\cot \theta=a$, see Fig.~\ref{fig:parapicture} below.

\begin{figure}[b]
\centering
\includegraphics[scale=1]{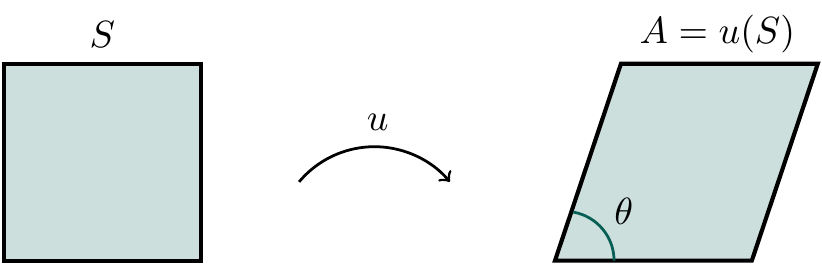}\qquad
\caption{Getting a parallelogram from a square through a simple volume preserving linear transformation.}
\label{fig:parapicture}
\end{figure}

In this case we obtain
\begin{equation}
    \label{eq:para}
    \mathcal{G}_{u(S)}=(L-M[x_1-ay_1,\ldots,x_n-a y_n])(L-M[y_1,\ldots,y_n])\,,
\end{equation}
provided the right-hand side is positive (otherwise the result is zero). This result will play a key role in Appendix~\ref{sec:divergence}. Notice another difference with the disc: for a parallelogram, the smallest order in the expansion is $L^0$, which means the asymptotic expansion terminates at order $L^0$. The constant term can be identified as a sum of corner terms, which are absent in the disc expansion. 

To be more precise, if we denote by $-d_m(\theta)$ the constant term in the expansion of the $m-$th cumulant $C_m(A)$ on the parallelogram, we have the exact formula, using complex numbers instead of two-dimensional vectors:
\begin{equation}\label{eq:generalcornerpara}
    d_m(\theta)=-\int dx_2\ldots dx_m dy_2\ldots dy_m M[x_2-a y_2,\ldots,x_m-ay_m]M[y_2,\ldots,y_m] f(x_2+i y_2,\ldots,x_m+i y_m).
\end{equation}
All integrals are over $\mathbb{R}$. The somewhat artificial minus signs were introduced to match with existing literature on corner terms. We emphasize again that (\ref{eq:generalcornerpara}) is very general, and in particular it holds also for interacting systems. In terms of single corner functions $a_m(\theta)$, we have $d_m(\theta)=2a_m(\theta)+2 a_m(\pi-\theta)$, so it provides important information on the contribution of a single corner, but does not allow to fully reconstruct it.
\subsection{$1/\theta$ divergence for a single corner}
\label{sec:divergence}
The parallelogram results allow to extract the small angle behavior of the corner term, since $a_m(\pi-\theta)\to 0$ as  $\theta\to 0$. Using the estimate 
\begin{align}
    M[x_2-a y_2,\ldots,x_m-a y_m]&\sim a M[y_2,\ldots,y_m]\\
    &\sim \frac{1}{\theta} M[y_2,\ldots,y_m]
\end{align}
as $a\to\infty$ or $\theta\to 0$, it is easy to show that the corner term diverges as $1/\theta$ as $\theta\to 0$ in considerable generality. Recalling that $d_m(\theta)$ corresponds to the contribution of two corners with infinitesimal angles we arrive at
\begin{equation}
    a_m(\theta)\sim \frac{\kappa_m}{\theta}
\end{equation}
with 
\begin{equation}\label{eq:generalcornerdivergence}
    \kappa_m=-\frac{1}{2}\int dx_2\ldots dx_m dy_2\ldots dy_m M[y_2,\ldots,y_m]^2 f(x_2+i y_2,\ldots,x_m+i y_m).
\end{equation}

\section{Some exact results for corner terms in the integer quantum Hall effect}
\label{sec:smallangles}

In this appendix, we apply the general results of Appendix~\ref{sec:ultragenereral} to the integer quantum Hall effect, which is a bulk 2D free fermions system with two--point function (or kernel) given in symmetric gauge by
\begin{equation}\label{eq:bulkiqh}
    K(z,w)=\frac{1}{\pi} e^{-\frac{1}{2}(|z-w|^2-z^*w+w^* z)}.
\end{equation}
One way to use our general results is to reconstruct the connected $m$--point density function using Wick's theorem. While this is certainly doable, we find it more convenient to use the well known result \cite{Klich:2008un}
\begin{align}
     C_m&=\textrm{Tr} \,\partial_\lambda^m     \left.\log\left[1+(e^\lambda-1)K\right]\right|_{\lambda=0}\\
     &=\sum_{p=1}^m \beta_{pm}\textrm{Tr}\, K^p \,, 
\end{align}
for known coefficients $\beta_{pm}=\frac{1}{p}\sum_{j=0}^p (-1)^{j+1}C_p^j j^m$, 
which implies that $C_m$ is the trace of a polynomial\footnote{\mbox{The first few  are $C_2=\textrm{Tr} (K-K^2)$, $C_3=\textrm{Tr} (K-3K^2+2K^3)$, $C_4=\textrm{Tr}(K-7K^2+12K^3-6K^4)$, etc.}} of degree $m$ in $K$, and one can compute each trace $\textrm{Tr}\, K^p$ separately. Our main expansion result can then be applied because $K(z_1,z_2)\ldots K(z_{m-1},z_m)K(z_m,z_1)$ is indeed translational invariant, so we may use our main formula (\ref{eq:funda}) and several others with the left-hand side replaced by $\textrm{Tr}\, K^m$, and $f$ in the right-hand side replaced by
\begin{equation}\label{eq:relevantf}
    g(z_2,\ldots,z_m)=K(0,z_2)K(z_2,z_3)\ldots K(z_{m-1},z_m)K(z_m,0)\,.
\end{equation}

\subsection{Corner terms in a parallelogram}
 In this subsection, $A$ is again the image of the square $S=[0,L]^2$ through the map $u((x,y))=(x+ay,y)$, that is, $A=u(S)$ is a parallelogram with four angles $\theta,\pi-\theta,\theta,\pi-\theta$, where $\cot \theta=a$. 
Using equations \eqref{eq:funda} and \eqref{eq:para}, $\textrm{Tr}\, K^m$ has full asymptotic expansion
\bea
    \label{eq:fulltrkm}
   &\textrm{Tr}\, K^m=\nn\\
   &\int d x_2\ldots dx_m dy_2\ldots dy_m \, (L-M[x_2-ay_2,\ldots,x_m-a y_m])(L-M[y_2,\ldots,y_m]) g(x_2+iy_2,\ldots,x_m+i y_m)+O(L^{-\infty})\,,
\eea
\pagebreak
where $f$ is given by (\ref{eq:bulkiqh}) and (\ref{eq:relevantf}), and all integrals are over $\mathbb{R}$. The constant term in this asymptotic expansion is
\begin{equation}
    k_m(\theta)=\int d x_2\ldots dx_m dy_2\ldots dy_m  M[x_2-ay_2,\ldots,x_m-a y_m] M[y_2,\ldots,y_m] g(x_2+iy_2,\ldots,x_m+i y_m)
\end{equation}
for $m\geq 2$, and $k_1(\theta)=0$. 
Therefore, (minus) the constant contribution to the $m$--th cumulant on the parallelogram $u(S)$ may be reconstructed as
\begin{equation}
    d_m(\theta)=-\sum_{p=1}^m \beta_{pm} k_m(\theta)\,.
\end{equation}
Recall that in terms of the single corner function, $d_m$ reads
\begin{equation}
    d_m(\theta)=2a_m(\theta)+2 a_m(\pi-\theta),
\end{equation}
Therefore, the above analytical result for $d_m(\theta)$ is not sufficient to fully reconstruct $a_m(\theta)$ in general. In Appendix~\ref{sec:exactnumerics}, we demonstrate how this may be circumvented by studying IQH on a cylinder geometry. Using this method, one can obtain numerical estimates for all $a_m$, essentially to arbitrary precision. However, fully analytical calculations seem to be difficult with this approach.  

There are nevertheless two interesting cases where the limitations of the parallelogram results can be avoided. The first is the regime $\theta\to 0$, as already done in Appendix~\ref{sec:divergence}. The other one is $\theta=\pi/2$, in which case $d_m(\pi/2)=4a_m(\pi/2)$ (Appendix~\ref{sec:rightcorner}).

\subsection{$1/\theta$ divergence for a single corner}

Let us denote by $\bar{k}_m$ the coefficient of the divergence of $k_m(\theta)$ as $\theta\to 0$. By the reasoning leading to  (\ref{eq:generalcornerdivergence}), we obtain 
\begin{align}
    \bar{k}_m&=\int_{\mathbb{R}^{2m-2}}dx_2,\ldots dx_m dy_2\ldots dy_m\,  M[y_2,\ldots,y_m]^2 g(x_2+i y_2,\ldots,x_m+i y_m)\\\label{eq:forbound}
    &=\frac{1}{\sqrt{2\pi m}}\left(\frac{2}{\pi}\right)^{\frac{m}{2}}\int_{\mathbb{R}^{m-1}}dy_2 \ldots dy_m\, M[y_2,\ldots,y_m]^2\exp\left[-2\sum_{j=2}^m y_j^2+\frac{2}{m}\left(\sum_{j=2}^m y_j\right)^2\right]\\
    &=\frac{(m-1)!}{2\sqrt{\pi m}\,\pi^{\frac{m}{2}}}\int_{y_2<\ldots <y_m}dy_2 \ldots dy_m \big(\max(0,y_m)-\min(0,y_2)\big)^2\exp\left[-\sum_{j=2}^m y_j^2+\frac{1}{m}\left(\sum_{j=2}^m y_j\right)^2\right].
\end{align}
The above can be simplified to
\begin{equation}
    \bar{k}_m=\frac{m!}{2\sqrt{\pi m}\,\pi^{\frac{m}{2}}}\int_{0<y_2<\ldots <y_m}dy_2 \ldots dy_m\, y_m^2\exp\left[-\sum_{j=2}^m y_j^2+\frac{1}{m}\left(\sum_{j=2}^m y_j\right)^2\right]
\end{equation}
after further manipulations. We managed to compute the integrals analytically for $m\in\{2,3,4\}$ and numerically otherwise. The result can be used to determine the coefficient of the corner divergence $\kappa_m$ for the $m$--th cumulant, once again by expressing the cumulant in terms of traces of powers of $K$. We obtain in particular
\begin{align}
    \kappa_2&=\frac{\bar{k}_2}{2}=\frac{1}{4\pi}\simeq 0.07957747155\,,\\
    \kappa_3&=-\frac{-3\bar{k}_2+2\bar{k}_3}{2}=-\frac{3\sqrt{3}-\pi}{4\pi^2}\simeq -0.05204260692\,,\\
    \kappa_4&=-\frac{-7\bar{k}_2+12\bar{k}_3-6\bar{k}_4}{2}=\frac{-12\sqrt{3}+\pi+18}{4\pi^2}\simeq 0.009042484082\,,\\
    \kappa_5&=-\frac{-15 \bar{k}_2+50 \bar{k}_3-60 \bar{k}_4+24 \bar{k}_5}{2}\simeq 0.01168589257\,,
\end{align}
and so on. Those are in perfect agreement with the numerically exact results of the main text, obtained using the method described in Appendix~\ref{sec:exactnumerics}. 
\subsection{$\theta=\pi/2$ for a single corner}
\label{sec:rightcorner}
This case simply corresponds to the square $a=0$. We obtain
\begin{equation}
    k_m(\pi/2)=\int_{\mathbb{R}^{2m-2}}dx_2\ldots dx_m dy_2\ldots dy_m M[x_2,\ldots,x_m]M[y_2,\ldots,y_m]g(x_2+iy_2,\ldots,x_m+iy_m)\,.
\end{equation}
One gets
\begin{equation}
    k_2(\pi/2)=\frac{1}{\pi^2} \,, \qquad
    k_3(\pi/2)=\frac{3(1+\sqrt{5})+\log(\sqrt{5}-2)}{4\pi^2}\,,
\end{equation}
after a very long calculation for $m=3$. In terms of pure corner terms for the cumulants, we obtain
\begin{align}
a_2(\pi/2)&=\frac{k_2(\pi/2)}{4}=\frac{1}{4\pi^2}\simeq 0.02533029591 \,,\\
a_3(\pi/2)&=-\frac{-3k_2(\pi/2)+2k_3(\pi/2)}{4}=-\frac{3(\sqrt{5}-1)+\log(\sqrt{5}-2)}{8\pi^2}\simeq -0.0286810945\,,
\end{align}
where recall the factor $1/4$ accounts for the fact that there are four corners with angle $\pi/2$ in the square.

\subsection{Breakdown of superuniversality for cumulants higher than variance}

Consider the ratios $\kappa_m/a_m(\pi/2)$ which compare the behavior at $\theta=\pi/2$ and  $\theta\to 0$. For IQH states, our previous results imply
\begin{align}
    \frac{\kappa_2}{a_2(\pi/2)}&=\pi \,,\\
    \frac{\kappa_3}{a_3(\pi/2)}&=\frac{6 \sqrt{3}-2 \pi }{3 \sqrt{5}-3+\log (\sqrt{5}-2)} \simeq 1.814526527\,.
\end{align}
The ratio for the second cumulant is always $\pi$ for any theory provided $f$ does not decay too slowly, due to superuniversality \cite{Estienne:2021hbx}. There is no reason why this would hold for higher cumulants. A counter-example is provided by a free fermion theory with pure gaussian kernel
\begin{equation}
    V(z,w)=\frac{1}{\pi} e^{-|z-w|^2/2}\,,
\end{equation}
which is similar to the IQH kernel (\ref{eq:bulkiqh}), but simpler for cumulants higher than the variance. Using similar techniques as those described above we get
\begin{align}
    \frac{\kappa_2}{a_2(\pi/2)}&=\pi\,,\\
    \frac{\kappa_3}{a_3(\pi/2)}&=\frac{48\sqrt{3}+5\pi}{45} \simeq 2.196586712\,.
\end{align}
Hence the ratio for the third cumulant differs from that of IQH, demonstrating a breakdown of superuniversality. From a technical standpoint, the pure gaussian kernel is significantly simpler than its IQH counterpart at angle $\pi/2$, because it is translationally invariant\footnote{\mbox{If $V(z,w)=V(z-w)$, then $V(z_1+w,z_2+w)\ldots V(z_m+w,z_1+w)=V(z_1,z_2)\ldots V(z_m,z_1)$, but the converse is not true, a counter-} \mbox{example being provided precisely by IQH. Why translation kernel are special is nicely explained  in \cite{Kac1954Toeplitz}.}}. In this case it is possible to exploit the results of \cite{Kac1954Toeplitz} even further, and get the formula
\begin{equation}
    k_m(\pi/2)=\left[\frac{2^{m/2-1}}{\pi}\sum_{p=1}^{m-1}\frac{1}{\sqrt{p(m-p)}}\right]^2,
\end{equation}
from which one can reconstruct all $a_m(\pi/2)$. Note that for $m$ very large, $k_m(\pi/2)\simeq 2^{m-2}$.

\section{Cumulants for IQH states on a cylinder from the overlap matrix}
\label{sec:exactnumerics}

The single-electron Hamiltonian in the Landau gauge for IQH states is given by
\bea
H=\frac{p_x^2}{2m_e}+\frac{(p_y+e Bx)^2}{2m_e}\,,
\eea
where $m_e$ is the effective mass of the electron. The orientation is chosen so that $e B >0$, and we rescale $x$ and $y$ to set the magnetic length $\ell_B = \sqrt{\hbar/e B}$ to unity. On a two-dimensional cylinder of circumference $l_y$, the eigenstates of $H$ are organized into the Landau level (LL) labelled by $n\in \mathbb{N}$, with wavefunctions
\bea
\phi_{n,k}(x,y) = \frac{ e^{i k y}}{\sqrt{2^n n! \ell_y\sqrt{\pi}}}H_n\leftt x+k \rightt e^{-(x+k)^2/2 } , \qquad k \in 2\pi \mathbb{Z}/\ell_y\,,
\eea
where $H_n(x)$ are the Hermite polynomials. The many-body IQH state at filling fraction $\nu \in \mathbb{N}^*$ is obtained by entirely filling all LLs with $n < \nu$.

\begin{figure}[b]
\centering
\includegraphics[scale=0.9]{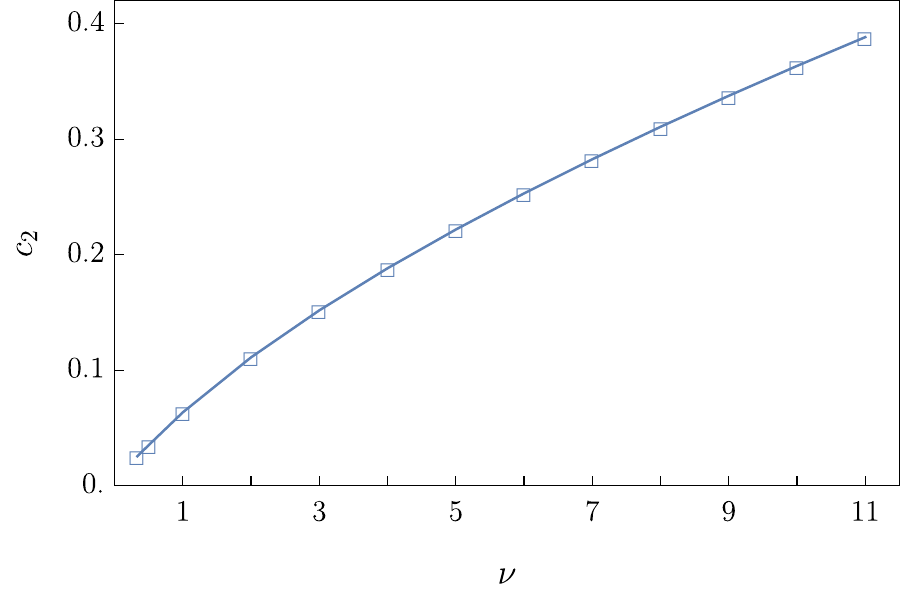}\hfill
\includegraphics[scale=0.862]{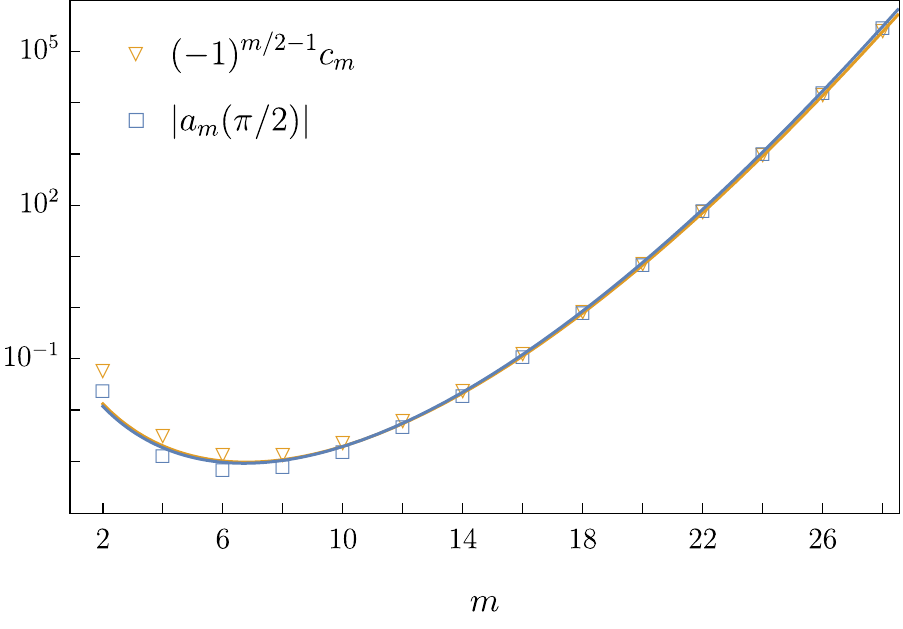}
\vspace{-5pt}
\caption{\textit{Left:} Area-law coefficient $c_2$ for the variance as a function of integer filling $\nu$. \textit{Right:} Area-law coefficient $c_m$ (orange) and corner cumulant function $|a_m(\pi/2)|$ (blue) for $m=2,4,\cdots,28$. The data ranges over 8 orders of magnitude (log-linear plot). The two solid lines show an ansatz of the form $d_1 (m-1)!/d_2^m$, where $d_1,\, d_2$ are fitting parameters.}
\lb{SM:fig1}
\end{figure}

Bipartite cumulants $C_m(A)$ for a subregion $A$ can be written in terms of the overlap matrix \cite{Klich:2004pb,2009PhRvB..80o3303R}
\bea\lb{overlap}
\mathbb{A}_{(n,k)\,(n',k')} = \int_A d^2{\bf r}\, \phi_{n,k}({\bf r})\phi_{n',k'}^*({\bf r})\,,
\eea
with $0\leq n,n' \leq \nu -1$ and $k, k'\in 2\pi \mathbb{Z}/\ell_y$. For each pair of momenta $(k,k')$ one has a block corresponding to the occupied LLs. The relation between cumulants and the overlap matrix is the following
\bes
&C_m(A)= \p_\lambda^m\log \chi_A(\lambda)|_{\lambda=0}\,,  \\ &\chi_A(\lambda)=\det[1+(e^{\lambda}-1)\mathbb{A}]\,,
\ees
where $\chi_A(\lambda)$ is the cumulant generating function. We give the explicit expressions for the first few cumulants:
\bes
&C_2=\tr\hspace{-1.5pt}\left(\mathbb{A}- \mathbb{A}^2 \right),\\ &C_3=\tr\hspace{-1.5pt}\left(\mathbb{A}- 3\mathbb{A}^2 +2\mathbb{A}^3 \right),\\
&C_4=\tr\hspace{-1.5pt}\left(\mathbb{A}- 7\mathbb{A}^2 +12\mathbb{A}^3 -6\mathbb{A}^4 \right),\\
&C_5=\tr\hspace{-1.5pt}\left(\mathbb{A}- 15\mathbb{A}^2 +50\mathbb{A}^3 -60\mathbb{A}^4 +24\mathbb{A}^5\right).
\ees

In the main text, we focus on the IQH state with filling fraction $\nu=1$ (setting $n=n'=0$ in \eqref{overlap}), and regions presenting corners. The strategy to extract the corner contribution in the even cumulants is the following.
We first compute the cumulants with the method described above for arrow-shaped regions as depicted in Fig.\,\ref{SM:fig_geo}, and then subtract the leading area law. The area-law coefficients are known explicitly \cite{2019CMaPh.376..521C}, i.e.
\bes\lb{eq:fcs_area}
c_m=\frac{1}{2\pi}\p_\lambda^m \left.\int_{-\infty}^{\infty}dk \,\log\hspace{-1pt}\big(1+(e^{\lambda}-1)\nu(k) \big)\right|_{\lambda =0}\,, \qquad\quad \nu(k)=\frac12(1-{\rm erf}(k))\,.
\ees 

\begin{figure}[t]
\centering
\raisebox{1.5cm}{\includegraphics[scale=0.9]{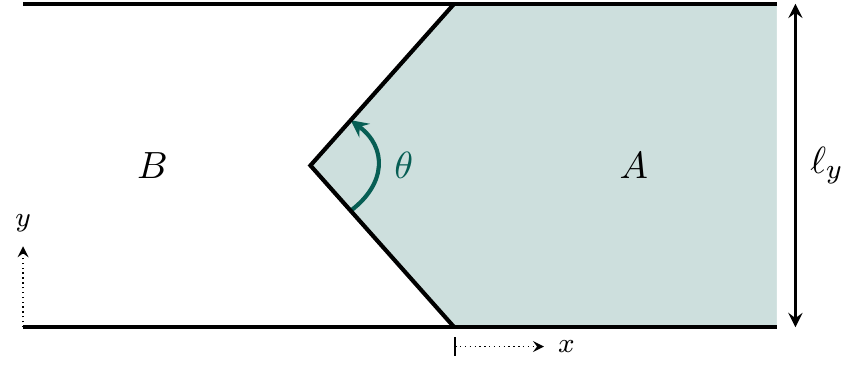}}\hfill
\includegraphics[scale=.97]{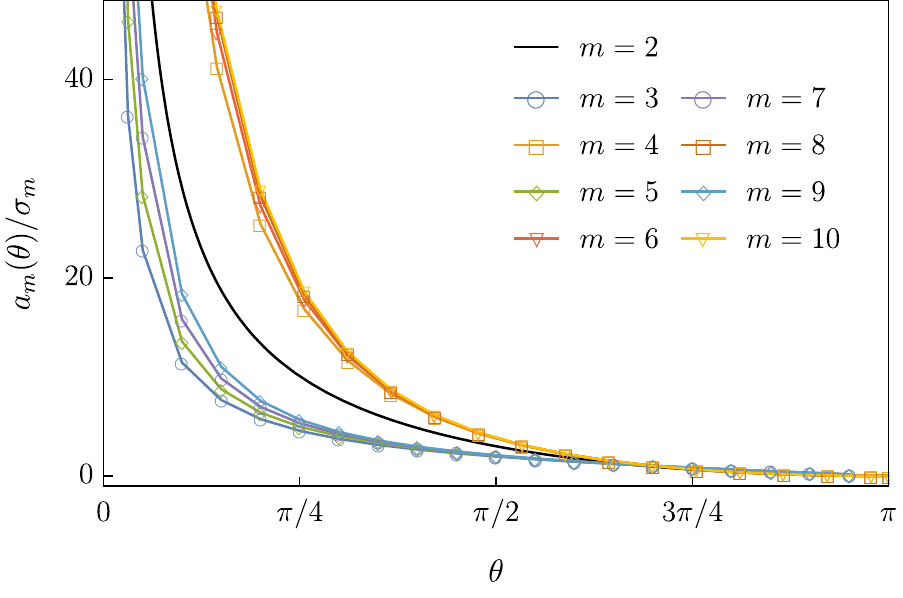}\vspace{-5pt}
\caption{\textit{Left:} Arrow-shaped region on the infinite cylinder used to extract the corner terms for even cumulants for IQH states. \textit{Right:} Normalized corner cumulant function $a_m(\theta)/\sigma_m$ for $m=2,3,4,\cdots,10$, for IQH $\nu=1$.}
\lb{SM:fig_geo}\vspace{-3pt}
\end{figure}

\noindent Finally, since such an arrow-shaped region possesses two corners of opening angles $\theta$ and $2\pi-\theta$, we use the symmetry for even cumulants $a_m(\theta)=a_m(2\pi-\theta)$ and divide our result by two to get $a_m(\theta)$.

To obtain the corner contribution for odd cumulants, one must work a little more. Indeed, because of the antisymmetry of odd cumulants for conserved charges, which translates for corners as $a_m(\theta)=-a_m(2\pi-\theta)$, we cannot use arrow-shaped regions to extract $a_m(\theta)$. Instead, we use combinations of different geometries. For example, starting with a square, we obtain $a_m(\pi/2)$. Next, we consider an isosceles right triangle, for which the corners contribution reads $a_m(\pi/2)+2a_m(\pi/4)$. We deduce $a_m(\pi/4)$ by subtracting the contribution of the angle $\pi/2$ previously obtained from the square. We may play this game for different shapes.

We present in Fig.\,\ref{SM:fig_geo} the normalized corner cumulants functions  $a_m(\theta)/\sigma_m$ for $m=2,3,4,\cdots,10$. One clearly recognizes that the variance stands out. The data of $a_3(\theta)$ and $a_4(\theta)$ for IQH and FQH can be found in Table \ref{tab3}.\vspace{-15pt}

\begin{table}[b]
\caption{\label{tab3}Cumulant corner function $a_m(\theta)$, $m=3,4$, for IQH $\nu=1$ state (exact), and FQH at $\nu=1/2,\,1/3$ (MC).}
\begin{ruledtabular}
\begin{tabular}{ c c c c c c c }
& \multicolumn{3}{c}{$a_3(\theta)$} & \multicolumn{3}{c}{$a_4(\theta)$}\vspace{3pt}\\
\cline{2-4} \cline{5-7}
$\theta$ & $\nu=1$ & $\nu=1/2$ & $\nu=1/3$ & $\nu=1$ & $\nu=1/2$ & $\nu=1/3$ \vspace{3pt}\\
\colrule
$\phantom{10}\pi/32$ & $-0.5310420027$ & --- & --- & $0.1017086853$ & --- & --- \\
$\phantom{10}\pi/16$ & $-0.2665374236$ & --- & --- & $0.0426363917$ & --- & --- \\
$\phantom{9}2\pi/20$ & $-0.1673981982$ & $-0.11053$ & $-0.08836$ & $0.0244449664$ & $0.0169$ & $0.01332$ \\
$\phantom{9}3\pi/20$ & $-0.1120365260$ & $-0.07484$ & $-0.05880$ & $0.0142661120$ & $0.0108$ & $0.00911$ \\
$\phantom{9}4\pi/20$ & $-0.0839258568$ & $-0.05617$ & $-0.04479$ & $0.00927085369$ & $0.00672$ & $0.00565$ \\
$\phantom{9}5\pi/20$ & $-0.0666408163$ & $-0.04466$ & $-0.03559$ & $0.00638696553$ & $0.00445$ & $0.00343$ \\
$\phantom{9}6\pi/20$ & $-0.0547456733$ & $-0.03673$ & $-0.02933$ & $0.00455976979$ & $0.00311$ & $0.00229$ \\
$\phantom{9}7\pi/20$ & $-0.0459251086$ & $-0.03081$ & $-0.02463$ & $0.00332933045$ & $0.00218$ & $0.00160$ \\
$\phantom{9}8\pi/20$ & $-0.0390255227$ & $-0.02618$ & $-0.02094$ & $0.00246419079$ & $0.00166$ & $0.00111$ \\
$\phantom{9}9\pi/20$ & $-0.0334061260$ & $-0.02242$ & $-0.01794$ & $0.00183612946$ & $0.00125$ & $0.000849$ \\
$10\pi/20$ & $-0.0286810945$ & $-0.01923$ & $-0.01536$ & $0.00136904036$ & $0.00093$ & $0.000623$ \\
$11\pi/20$ & $-0.0246030765$ & $-0.01652$ & $-0.01318$ & $0.00101534642$ & $0.00076$ & $0.000477$ \\
$12\pi/20$ & $-0.0210052160$ & $-0.01408$ & $-0.01123$ & $0.000744089820$ & $0.00057$ & $0.000341$ \\
$13\pi/20$ & $-0.0177700173$ & $-0.01190$ & $-0.00953$ & $0.000534487955$ & $0.00042$ & $0.000225$ \\
$14\pi/20$ & $-0.0148115522$ & $-0.00992$ & $-0.00794$ & $0.000372252397$ & $0.00032$ & $0.000157$ \\
$15\pi/20$ & $-0.0120647447$ & $-0.00807$ & $-0.00643$ & $0.000247393390$ & $0.00019$ & $0.000113$ \\
$16\pi/20$ & $-0.00947861342$ & $-0.00633$ & $-0.00509$ & $0.000152865323$ & --- & --- \\
$17\pi/20$ & $-0.00701181942$ & $-0.00469$ & $-0.00374$ & $0.0000837100448$ & --- & --- \\
$18\pi/20$ & $-0.00462960357$ & $-0.00311$ & $-0.00249$ & $0.0000365075225$ & --- & --- \\
$19\pi/20$ & $-0.00230157227$ & $-0.00154$ & $-0.00124$ & $0.00000902481865$ & --- & --- \\
$31\pi/32$ & $-0.00143681734$ & --- & --- & $0.00000351730833$ & --- & --- \\
\end{tabular}
\end{ruledtabular}
\end{table}


\newpage

\section{Cumulants of massless Dirac fermions}
\label{sec:dirac}

Consider a two-dimensional square lattice, infinite in one direction, say $x$, and impose antiperiodic boundary conditions in the other, $y$. We want to compute the charge cumulants of a section $A$ of the infinite cylinder, i.e. $A$ is a finite cylinder of length $\ell_x$ and circumference $\ell$. We may take advantage of the symmetry and perform a dimensional reduction along the transverse direction $y$.
The lattice Hamiltonian of a 2D free massless Dirac fermion reads
\bes
H = -\frac{i}{2}\sum_{i,j} \left[\Psi_{i,j}^\dagger\gamma^0\gamma^1(\Psi_{i+1,j}-\Psi_{i,j}) + \Psi_{i,j}^\dagger\gamma^0\gamma^2(\Psi_{i,j+1}-\Psi_{i,j}) - {\rm h.c.}  \right],\lb{Hf2d}
\ees
where we set the lattice spacing to unity.  The two-dimensional matrices $\gamma^0$ and $\gamma^j$ are proportional to Dirac matrices (e.g.\ $\gamma^0=\sigma_3$ and $\gamma^1=i\sigma_1,\gamma^2=i\sigma_2$, with $\sigma_{1,2,3}$ the Pauli matrices). 
After dimensional reduction along $y$ (indexed by $j$), the resulting Hamiltonian consists in a sum of decoupled 1D massive free Dirac fermions, $H=\sum_k H_k$,
\bes
H_k = \sum_i \left[-\frac{i}{2}\Big(\Psi_i^\dagger\gamma^0\gamma^1(\Psi_{i+1}-\Psi_i) - {\rm h.c.} \Big) + m_k\Psi_i^\dagger \gamma^0\gamma^2\Psi_i \right],\lb{Hf1d}
\ees
where $m_k=\sin k_y$ and $k_y=2\pi(y-1/2)/\ell$, with $\ell$ the length of the subregion $A$ along $y$. The eigenvalues of the reduced density matrix can be related to those of the correlation matrix $\la \Psi^\dagger\Psi\ra|_A$ restricted to a region $A$, see \cite{Chung:2001zz,2003JPhA...36L.205P}. 
The correlator for the 1D infinite chain is given by
\bes
\la \Psi_i^\dagger\Psi_{i'} \ra = \frac{1}2\delta_{ij}+ \frac{1}{4\pi}\int_{-\pi}^\pi dx \frac{\gamma^0\gamma^2m_k+\gamma^0\gamma^1\sin x}{\sqrt{m_k^2+\sin^2x}}e^{ix(i-i')}\,.
\ees
The expression for the 1D charge cumulants in terms of the eigenvalues $\nu_k$ of $\la \Psi^\dagger\Psi\ra|_A$ reads
\bes
C^{(1d)}_m= \sum_i\p_\lambda^m\log [1+(e^{\lambda}-1)\nu_k]\,,
\ees
Since we have performed a dimensional reduction, the charge cumulants are obtain by summing over the modes as $C_m(A)= \sum_{k}C^{(1d)}_m(k)$,
where $C^{(1d)}_m(k)$ is the cumulant for the $k^{\rm th}$ mode associated to $H_k$.
Note that due to the fermion doubling on the lattice, one has to divide the lattice results by 4 to get the charge cumulants corresponding to a Dirac field in the continuum limit.

Since the spectrum of the correlation matrix is symmetric around $1/2$, the odd cumulants vanish exactly, no matter the subregion $A$ one chooses, as expected from charge conjugation symmetry. In contrast, in the limit of large $\ell,\ell_x$ we find that the even cumulants satisfy an area law, $C_m(A)=c_m 2\ell+\cdots$, whose coefficients have the following signs: $+,+,-,+,+,-,\cdots$ for $m=2,4,6,8,10,12$, see Table~\ref{tabS2}.

\begin{table}[h]
\caption{\label{tabS2}Area-law coefficients $c_m$, $m=2,4,6,8,10,12$, for the free massless Dirac fermion.}
\begin{ruledtabular}
\begin{tabular}{ l c c c c c c }
& \raisebox{1.5pt}{$m=2$} & \raisebox{1.5pt}{$m=4$} & \raisebox{1.5pt}{$m=6$} & \raisebox{1.5pt}{$m=8$} & \raisebox{1.5pt}{$m=10$} & \raisebox{1.5pt}{$m=12$}\\
\colrule
$c_m$ & $0.020621681$ & $0.0084800$ & $-0.009303$ & $0.00471$ & $0.014$ & $-0.07$ \\
\end{tabular}
\end{ruledtabular}
\end{table}

\end{document}